\documentclass{article}

\usepackage{arxiv}

\usepackage[utf8]{inputenc} 
\usepackage[T1]{fontenc}    
\usepackage{hyperref}       
\usepackage{url}            
\usepackage{booktabs}       
\usepackage{amsfonts}       
\usepackage{nicefrac}       
\usepackage{microtype}      
\usepackage{graphicx}
\usepackage{amsmath}
\usepackage{amssymb}
\usepackage{bm}
\usepackage{algorithm}
\usepackage{algpseudocode}

\DeclareMathOperator*{\argmin}{argmin}
\title{\textit{In Vitro} Electron Density Refinement from Solution X-ray Scattering in the Wide-Angle Regime}

\author{Yen-Lin Chen \\
	School of Applied and Engineering Physics\\
	Cornell University\\
	Ithaca, NY 14853 \\
	\texttt{yc2253@cornell.edu} \\
	\And
	Lois Pollack \\
	School of Applied and Engineering Physics\\
	Cornell University\\
	Ithaca, NY 14853 \\
	\texttt{lp26@cornell.edu} \\}

\renewcommand{\headeright}{}
\renewcommand{\undertitle}{}
\newcommand{\angstrom}{\textup{\AA}}

\begin{document}
\maketitle


\begin{abstract}
	We present \textit{Frequency Marching}, \textit{FM}, an algorithm that refines three-dimensional electron density distributions from solution X-ray scattering data in both the small- and wide-angle regimes. This algorithm is based on a series of optimization steps, marching along the frequency (reciprocal) space and refining detailed periodic structures with the corresponding real-space resolution. Buffer subtraction and excluded volumes, key factors in extracting the signatures of the biomolecule of interest from the sample, are accounted for using implicit density models. We provide the numerical and analytical basis of the \textit{FM} algorithm. We demonstrate this technique by application to structured and unstructured nucleic acid systems, where higher resolution features are carved out of low resolution reconstructions as the algorithm marches into wider angles.
\end{abstract}


\section{Introduction}

The key to understanding how biomolecules function can often be found in their structures. The use of X-rays has greatly advanced our knowledge of molecular structure, primarily through X-ray crystallography, in which the signal from many identical molecules is interfered to create a signature that reflects a unique structure, often with atomic resolution. However, many molecules, especially those that are very dynamic, or  exist in a wide variety of states, resist the preparation required for crystallographic study. For these systems, solution X-ray scattering can be an important tool in determining the \textit{in vitro} structure of biomolecules. Because its only requirement is solubility, it serves as an essential complement to other structural techniques, such as X-ray crystallography, or even cryogenic electron microscopy (cryo EM), where complex strategies must sometimes be employed to ensure compatibility of sample with method. 

At the most fundamental level, solution scattering profiles reflect the three-dimensional excess electron density of biomolecules relative to the solvent that surrounds them (solvent background). The measured scattering intensity is most commonly integrated at a fixed angle relative to the direction of the incident X-ray beam, and converted into a 1D scattering profile showing the intensity of scattered X-rays at an angle $2\theta$ relative to the incident beam versus momentum transfer, or $q$. The momentum transfer is defined as $q = (4\pi/\lambda)\sin(\theta)$, where $\lambda$ is the incident X-ray wavelength and $\theta$ is half of the scattering angle. The real-space resolution of solution scattering measurements is $2\pi/q_{max}$ where $q_{max}$ is the largest scattering angle detected. 

Small-angle X-ray scattering (SAXS) typically considers scattering angles up to 3 degrees, about $q=0.3\angstrom^{-1}$, providing spatial discrimination on the order of tens of $\angstrom$. Even at this low resolution, SAXS reveals important molecular features, such as the overall size and shape of the biomolecules as radius of gyration ($R_g$) and pair-distance distribution ($P(r)$) between all electrons in the sample. The reconstruction of real-space three-dimensional molecular envelopes\cite{Franke2009} or electron densities\cite{Grant2018} can be used to visualize the data in real, as opposed to reciprocal space. In other important applications, SAXS is used to refine atomic-resolution models generated from chain building\cite{Plumridge2016}, molecular relaxation\cite{Mauney2018} and molecular dynamics (MD) simulations\cite{Plumridge2018, Chen2019} to revealing conformational ensembles \textit{in vitro}.

Wide-angle X-ray scattering, or WAXS extends solution scattering to higher angles, reaching $q\approx 1.0\angstrom^{-1}$ and beyond. WAXS is less employed than SAXS despite the obvious benefit of higher spatial resolution. Recent advances in beamline design, coupled with the availability of highly sensitive area detectors\cite{DiFabio2016, Yang2020} permit routine acquisition of robust WAXS profiles with high signal-to-noise ratios ($S/N$). WAXS has already provided new insights into protein crowding effects\cite{Makowski2008}, solution ensembles\cite{Chen2014}, periodic nucleic acid structures\cite{Pabit2016, Chen2019b} and solvent models\cite{Virtanen2011, Nguyen2014}.  The information content in WAXS profiles can be dramatically enhanced when combined with atomic models for quantitative interpretation or to discriminate among constructs for qualitative comparisons. Structural parameters can be readily extracted from models using statistical modeling tools\cite{Chen2020}. However, all of the above approaches require atomic models from simulations, which can be challenging to perform for some macromolecular systems. A natural question arises: what can we learn from WAXS profiles in the absence of atomic models?

We propose a new algorithm called \textit{Frequency Marching}. \textit{FM} is a step-wise optimization algorithm that refines the 3D electron density using the full spectrum of solution X-ray scattering data, both SAXS and WAXS. The electron density is modeled using 3D mesh-grids or voxels whose size corresponds to the resolution and therefore, is related to range of scattering angles $q$. The buffer or solvent background, which must be removed, leaving only the scattering from the biomolecule, is modeled as a uniform and constant electron density on all the voxels. Buffer subtraction and excluded volumes are accounted for in the computation of solution X-ray scattering profile. \textit{FM} is implemented using a series of optimization steps with increasing scattering angle, $q$. Beginning with a molecular envelope or low resolution shape reconstructed from low angle data (large spatial dimension, the first step of \textit{FM} is to refine the electron density within the envelope against the experimental data from $q_{min}$ up to $q=q^{(1)}$, where $q_{min}\approx 0$ is the smallest angle achieved. We refer to this process as \textit{carving} the electron density. The resulting \textit{carved} density is subsequently used as the starting point for second round of \textit{carving} process corresponding to higher $q=q^{(2)}>q^{(1)}$ and so on. This process repeats as higher q data are appended, effectively reducing the size of 3D voxels, until the resolution corresponding to the specified $q_{max}$ is reached. Density from solvent or ion shells appears automatically during the \textit{carving} process where electron-dense structures appear. \textit{FM} enables the interpretation of solution X-ray scattering data in the wide-angle regime when atomic models are absent. It \textit{carves} out fine structural features that are beyond the resolution of SAXS domain and exhibits its greatest potential in the study of molecules with repeated structures, periodic molecules in solution such as nucleic acids, lipids, collagen and so on. 

The name of \textit{Frequency Marching} is derived from a step-wise linearization algorithm in solving cryoEM reconstruction\cite{Barnett2017}. In the next section, we will formulate the problem and provide numerical and statistical foundations. In the later sections, we will provide the pseudo-code and demonstration of \textit{FM} algorithm and its application to study ribonucleic acid (RNA) systems. Because of the repeated length scales that appear in RNA structures, signaling fundamental building blocks such as duplexes or triplexes, \textit{FM} is especially informative in reporting the structures of these important biomolecules.

\section{Problem Formulation and Analysis}

The algorithm begins with one FM step in $q$-space from $q_{min}$ to $q_{max}$, where $q_{min}\approx 0.01 \angstrom^{-1}$ and $q_{max}$ here is the largest scattering angle of this FM step, corresponding to the resolution. At this FM step, $n$ scattering angles are sampled: $q_{min} = q_1 < q_2 < \hdots < q_n = q_{max}$ and the goal is to minimize the customized $\chi^2$ of the following form.
\begin{equation}
\chi^2 = \frac{1}{n-1} \sum_{i=1}^n \left\{ \frac{\log_{10}\left[ I(q_i, \bm{d}) \right] - \log_{10} \left[ I_{exp}(q_i) \right] }{\sigma'(q_i)} \right\}^2,
\label{eq:1}
\end{equation}
where $I(q_i, \bm{d})$ is the computed scattering intensity from 3D electron density distribution $\bm{d} \in \mathbb{R}^N$ on the support of molecular shape.  $I_{exp}(q_i)$ is the experimentally measured scattering intensity and $\sigma'(q_i)$ is the propagated error at $q_i$.
\begin{equation}
\sigma'(q_i) = \frac{\sigma(q_i)}{I_{exp}(q_i)\log10} = \frac{1}{\left[ \frac{S}{N} \right]_i \log10}
\label{eq:2}
\end{equation}
In Eq. \eqref{eq:2}, $\sigma(q_i)$ and $\left[ \frac{S}{N} \right]_i$ are the experimental error and signal-to-noise ratio at $q_i$ respectively. Our problem at this FM step can be formulated as: 
\begin{equation}
\hat{\bm{d}} = \argmin_{\bm{d} \in \mathbb{R}^N} \chi^2 = \argmin_{\bm{d} \in \mathbb{R}^N} \frac{1}{n-1} \sum_{i=1}^n \left\{ \frac{\log_{10}\left[ I(q_i, \bm{d}) \right] - \log_{10} \left[ I_{exp}(q_i) \right] }{\sigma'(q_i)} \right\}^2
\label{eq:3}
\end{equation}

\subsection{The Gradient}

The optimizer, $\hat{\bm{d}}$, in Eq. \eqref{eq:3} is the refined 3D electron density. For many optimization algorithms, the gradient of the objective function is helpful in the implementation and can be used to analyze the numerical stability. We begin to calculate the gradient of $\chi^2$ defined in Eq. \eqref{eq:1}. 

\begin{equation}
\chi^2 = \frac{1}{n-1} \sum_{i=1}^n \chi_i^2
\end{equation}

\begin{equation}
\chi_i^2 = \left[ \frac{1}{\sigma'(q_i)} \right]^2 \Bigl\{ \log_{10}\left[ I(q_i, \bm{d}) \right] - \log_{10} \left[ I_{exp}(q_i) \right] \Bigr\}^2
\label{eq:5}
\end{equation}

\begin{equation}
\nabla_{\bm{d}} \left(\chi^2\right) = \frac{1}{n-1} \sum_{i=1}^n \nabla_{\bm{d}} \left(\chi_i^2\right)
\label{eq:6}
\end{equation}
\begin{equation}
\nabla_{\bm{d}} \left(\chi_i^2\right) = \frac{2}{I(q_i, \bm{d})} \left[ \frac{I_{exp}(q_i)}{\sigma(q_i)} \right]^2 \Bigl\{ \log\left[ I(q_i, \bm{d}) \right] - \log\left[ I_{exp}(q_i) \right] \Bigr\} \nabla_{\bm{d}} \left[ I(q_i, \bm{d}) \right] 
\label{eq:7}
\end{equation}
In order to proceed with $\nabla_{\bm{d}} \left[ I(q_i, \bm{d}) \right]$, we have to derive $I(q, \bm{d})$ from the computations of solution X-ray scattering profile using orientational average\cite{Park2009}. The buffer-subtracted solution X-ray scattering intensity is $I(q) = I_A(q) - I_B(q)$ where $I_A(q)$ and $I_B(q)$ are the scattering intensities from the solute and solvent respectively. Let $D(\bm{q})$ be the scattering intensity at a specific solid angle $\bm{q}$. We have 
\begin{equation}
I(q) = \langle \left\vert A(\bm{q}) - B(\bm{q}) \right\vert^2 \rangle_\Omega = \langle D(\bm{q}) \rangle_\Omega,
\label{eq:8}
\end{equation}
where $\langle\quad\rangle_\Omega$ denotes the average over the 3D orientations. Moreover, 
\begin{equation}
\label{eq:9}
\begin{split}
D(\bm{q}) & = \left\vert A(\bm{q}) - B(\bm{q}) \right\vert^2 \\
 & = \left\vert A(\bm{q}) \right\vert^2 - \left\vert B(\bm{q}) \right\vert^2 + \left[ 2\left\vert B(\bm{q}) \right\vert^2 - A^*(\bm{q})B(\bm{q}) - B^*(\bm{q})A(\bm{q})\right]\\
 & = \left\vert A(\bm{q}) \right\vert^2 - \left\vert B(\bm{q}) \right\vert^2 - 2\Re\left\{B^*(\bm{q})\times \left[ A(\bm{q}) - B(\bm{q}) \right] \right\}
\end{split}
\end{equation}
In Eq. \eqref{eq:8} and \eqref{eq:9}, $A(\bm{q})$ and $B(\bm{q})$ are the complex scattering amplitudes of the solute and solvent systems respectively. 
\begin{equation}
\label{eq:10}
A(\bm{q}) = \sum_{k=1}^{N} f_k(q)e^{-i\bm{q}\cdot\bm{r_k}}
\end{equation}
where $f_k(q)$ is the scattering form factor (electron density) at one support voxel of the molecular shape with coordinate $\bm{r_k}$. Essentially, $f_k(q) = d_k$, where $d_k$ is the k\textsuperscript{th} element of the vector $\bm{d}$. $N$ is the number of voxels within the molecular shape. The same formulation applies to the solvent system $B(\bm{q})$ as well. With Eq. \eqref{eq:9} and \eqref{eq:10}, we can now proceed to derive the buffer-subtracted solution X-ray scattering intensity.

First, let $\bm{R}$ be the coordinate matrix of all the voxels within the molecular support. 
\begin{equation}
\bm{R} = \begin{bmatrix}
x_1 & y_1 & z_1 \\
x_2 & y_2 & z_2 \\
\vdots & \vdots & \vdots \\ 
x_N & y_N & z_N
\end{bmatrix}
\end{equation}
Let $\bm{Q}$ be the matrix containing $J$ uniformly distributed orientations that span the $4\pi$ solid angle\cite{Ponti1999}, i.e.
\begin{equation}
\bm{Q} = \begin{bmatrix}
\sin\theta_1\cos\phi_1 & \sin\theta_1\sin\phi_1 & \cos\theta_1 \\
\sin\theta_2\cos\phi_2 & \sin\theta_2\sin\phi_2 & \cos\theta_2 \\
\vdots & \vdots & \vdots \\
\sin\theta_J\cos\phi_J & \sin\theta_J\sin\phi_J & \cos\theta_J 
\end{bmatrix}
\end{equation}
Second, define two matrices $\bm{A}$ and $\bm{B}$ as the following
\begin{equation}
\begin{split}
\bm{A} & = \cos(q\bm{R}\bm{Q}^T)\\
\bm{B} & = \sin(q\bm{R}\bm{Q}^T)
\end{split}
\end{equation}
The functions $\cos()$ and $\sin()$ are element-wise applications. Therefore, $\bm{A}, \bm{B} \in \mathbb{R}^{N\times J}$. The solute matrix, $\bm{U}\in\mathbb{R}^{2\times J}$, whose k\textsuperscript{th} column contains the real and imaginary part of the scattering amplitude $A(\bm{q_k})$ at orientation $\bm{q_k}$, can be written as the following. 
\begin{equation}
\bm{U} = 
\begin{bmatrix}
\Re\left[ A(\bm{q_1})\right] & \Re\left[ A(\bm{q_2})\right] & \dots & \Re\left[ A(\bm{q_J})\right]\\
\Im\left[ A(\bm{q_1})\right] & \Im\left[ A(\bm{q_2})\right] & \dots & \Im\left[ A(\bm{q_J})\right]
\end{bmatrix}
 = 
\begin{bmatrix}
\bm{d}^T\bm{A}\\
-\bm{d}^T\bm{B}
\end{bmatrix}
\end{equation}
Similarly, the solvent matrix is
\begin{equation}
\bm{V} = 
\begin{bmatrix}
\bm{c}^T\bm{A}\\
-\bm{c}^T\bm{B}
\end{bmatrix},
\end{equation}
where vector $\bm{c}$ is the constant electron density of the solvent background. Moreover, we define the excessive electron density vector $\bm{v}$ as $\bm{v} = \bm{d} - \bm{c}$. The buffer-subtracted matrix of the scattering amplitude, $\bm{M}$, is 
\begin{equation}
\bm{M} = \bm{U} - \bm{V} = 
\begin{bmatrix}
\bm{v}^T \bm{a_1} & \bm{v}^T \bm{a_2} & \bm{v}^T \bm{a_3} & \hdots & \bm{v}^T \bm{a_J}\\
-\bm{v}^T \bm{b_1} & -\bm{v}^T \bm{b_2} & -\bm{v}^T \bm{b_3} & \hdots & -\bm{v}^T \bm{b_J}
\end{bmatrix},
\end{equation}
where $\bm{a_k}$ and $\bm{b_k}$ are columns of $\bm{A}$ and $\bm{B}$ at orientation $k$. Using the same notation, $\bm{m_k}$ is the k\textsuperscript{th} column of $\bm{M}$. The scattering intensity at orientation $k$, $D_k$ is then
\begin{equation}
D_k = \left\Vert \bm{m_k} \right\Vert^2 = \left( \bm{v}^T \bm{a_k} \right)^2 + \left( \bm{v}^T \bm{b_k} \right)^2
\end{equation}
The orientationally averaged scattering intensity at $q$ from electron densities $\bm{d}$, $I(q, \bm{d})$ is 
\begin{equation}
\label{eq:18}
I(q, \bm{d}) = \frac{1}{J}\sum_{k=1}^J D_k = \frac{1}{J}\sum_{k=1}^J \Bigl[ \left( \bm{v}^T \bm{a_k} \right)^2 + \left( \bm{v}^T \bm{b_k} \right)^2 \Bigr]
\end{equation}
Therefore, $\nabla_{\bm{d}} \left[ I(q, \bm{d}) \right]$ can be derived using Eq. \eqref{eq:18}. 
\begin{equation}
\nabla_{\bm{d}} \left[ I(q, \bm{d}) \right] = \frac{2}{J}\sum_{k=1}^J \Bigl[ \left(\bm{v}^T\bm{a_k} \right)\bm{a_k} + \left(\bm{v}^T\bm{b_k} \right)\bm{b_k} \Bigr]
\label{eq:19}
\end{equation}
Combining Eq. \eqref{eq:7} and Eq. \eqref{eq:19}, 
\begin{equation}
\begin{split}
\nabla_{\bm{d}} \left(\chi_i^2 \right) = \frac{4}{I(q_i, \bm{d}) J} \left[ \frac{I_{exp}(q_i)}{\sigma(q_i)} \right]^2 & \Bigl\{ \log\left[ I(q_i, \bm{d}) \right] - \log\left[ I_{exp}(q_i) \right] \Bigr\} \times \\ 
 & \sum_{k=1}^J \Bigl[ \left(\bm{v}^T\bm{a_k} \right)\bm{a_k} + \left(\bm{v}^T\bm{b_k} \right)\bm{b_k} \Bigr]
\end{split}
\label{eq:20}
\end{equation}
Finally, with Eq. \eqref{eq:6} and Eq. \eqref{eq:20}, we arrive at
\begin{equation}
\begin{split}
\nabla_{\bm{d}} \left(\chi^2 \right) = \frac{1}{n-1} \sum_{i=1}^n  \frac{4}{I(q_i, \bm{d}) J} \left[ \frac{I_{exp}(q_i)}{\sigma(q_i)} \right]^2 & \Bigl\{ \log\left[ I(q_i, \bm{d}) \right] - \log\left[ I_{exp}(q_i) \right] \Bigr\} \times \\
 & \sum_{k=1}^J \Bigl[ \left(\bm{v}^T\bm{a_k} \right)\bm{a_k} + \left(\bm{v}^T\bm{b_k} \right)\bm{b_k} \Bigr] 
\end{split}
\label{eq:21}
\end{equation}
\begin{flushright}
$\blacksquare$\\
\end{flushright}

The gradient of our objective function, $\chi^2$, exists and is analytical. The Hessian, $\bm{H}=\nabla_{\bm{d}}^2 \left(\chi^2 \right)$ is also analytical and will be derived in the next section. However, $\bm{H} \in \mathbb{R}^{N\times N}$ is too large to be computationally practical. The gradient in Eq. \eqref{eq:21} is numerically stable since $n-1$, $I(q_i, \bm{d})$ and $J$ in the denominator all have finite positive values. The term, $\left[ \frac{I_{exp}(q_i)}{\sigma(q_i)} \right]$ is the experimental signal-to-noise ratio at $q_i$, $\left[ \frac{S}{N} \right]_i$, which ranges from about $10^{-1}$ to $10^3$ empirically. The gradient only starts to vanish as $\log\left[ I(q_i, \bm{d}) \right]$ approaches $\log\left[ I_{exp}(q_i) \right]$, i.e. $I(q_i, \bm{d}) \approx I_{exp}(q_i)$. Note that the vectors $\bm{a_k}$ and $\bm{b_k}$ depend on $q_i$ but are fixed with respect to $\bm{d}$. This property makes the computation of $\nabla_{\bm{d}} \left(\chi^2 \right)$ parallelizable and the computation of $\nabla_{\bm{d}}\left(\chi^2 \right)$ only marginally increases the run-time on top of the $\chi^2$. The above formulation is implemented in \textit{Julia}\cite{Bezanson2017} and it takes about 5 seconds for $N\approx7000$, $n=111$ and $J=1200$ on a 8-core PC.

\subsection{The Hessian}

The Hessian, $\bm{H} = \nabla_{\bm{d}}^2\left(\chi^2\right)$, can be derived by applying $\nabla_{\bm{d}}$ to the gradient derived in the last section. Its dimension is usually too large to be numerically useful but we'll derive it here for completeness. We start from Eq. \eqref{eq:7} and apply $\nabla_{\bm{d}}$. 
\begin{equation}
\bm{H_i} = \nabla_{\bm{d}} \left\{ \frac{2}{I(q_i, \bm{d})} \left[ \frac{I_{exp}(q_i)}{\sigma(q_i)} \right]^2 \Bigl\{ \log\left[ I(q_i, \bm{d}) \right] - \log\left[ I_{exp}(q_i) \right] \Bigr\} \nabla_{\bm{d}} \left[ I(q_i, \bm{d}) \right]  \right\}
\end{equation}
The Hessian, $\bm{H_i}$, is composed of two parts: 
\begin{equation}
\begin{split}
\nabla_{\bm{d}}^2 \left(\chi_i^2 \right) & = \frac{2}{\left[ I(q_i, \bm{d}) \right]^2} \left[ \frac{I_{exp}(q_i)}{\sigma(q_i)} \right]^2 \Bigl\{ 1 - \log\left[ I(q_i, \bm{d}) \right] + \log\left[ I_{exp}(q_i) \right] \Bigr\} \Bigl\{ \nabla_{\bm{d}} \left[ I(q_i, \bm{d}) \right]\Bigr\}\Bigl\{ \nabla_{\bm{d}} \left[ I(q_i, \bm{d}) \right]\Bigr\}^T\\
 & + \frac{2}{I(q_i, \bm{d})} \left[ \frac{I_{exp}(q_i)}{\sigma(q_i)} \right]^2 \Bigl\{ \log\left[ I(q_i, \bm{d}) \right] - \log\left[ I_{exp}(q_i) \right] \Bigr\} \nabla_{\bm{d}}^2 \left[ I(q_i, \bm{d}) \right]
\end{split}
\label{eq:23}
\end{equation}
We have yet to derive $\nabla_{\bm{d}}^2 \left[ I(q_i, \bm{d}) \right]$ and $\Bigl\{ \nabla_{\bm{d}} \left[ I(q_i, \bm{d}) \right]\Bigr\}\Bigl\{ \nabla_{\bm{d}} \left[ I(q_i, \bm{d}) \right]\Bigr\}^T$ but they are straightforward yet tedious to compute. 
\begin{equation}
\nabla_{\bm{d}}^2 \left[ I(q_i, \bm{d}) \right] = \frac{2}{J}\sum_{k=1}^J \Bigl( \bm{a_k}\bm{a_k}^T + \bm{b_k}\bm{b_k}^T \Bigr)
\label{eq:24}
\end{equation}

\begin{equation}
\begin{split}
\Bigl\{ \nabla_{\bm{d}} \left[ I(q_i, \bm{d}) \right]\Bigr\} & \Bigl\{ \nabla_{\bm{d}} \left[ I(q_i, \bm{d}) \right]\Bigr\}^T\\
 & = \left\{ \frac{2}{J}\sum_{k=1}^J \Bigl[ \left(\bm{v}^T\bm{a_k} \right)\bm{a_k} + \left(\bm{v}^T\bm{b_k} \right)\bm{b_k} \Bigr] \right\}\left\{ \frac{2}{J}\sum_{k=1}^J \Bigl[ \left(\bm{v}^T\bm{a_k} \right)\bm{a_k} + \left(\bm{v}^T\bm{b_k} \right)\bm{b_k} \Bigr] \right\}^T\\
 & = \frac{4}{J^2} \sum_{k=1}^J \sum_{l=1}^J \Bigl[ \left(\bm{v}^T\bm{a_k} \right)\left(\bm{v}^T\bm{a_l} \right)\bm{a_k}\bm{a_l}^T + \left(\bm{v}^T\bm{a_k} \right)\left(\bm{v}^T\bm{b_l} \right)\bm{a_k}\bm{b_l}^T \\
 & + \left(\bm{v}^T\bm{b_k} \right)\left(\bm{v}^T\bm{a_l} \right)\bm{b_k}\bm{a_l}^T + \left(\bm{v}^T\bm{b_k} \right)\left(\bm{v}^T\bm{b_l} \right)\bm{b_k}\bm{b_l}^T \Bigr]
\end{split}
\label{eq:25}
\end{equation}
By plugging Eq. \eqref{eq:24} and Eq. \eqref{eq:25} into Eq. \eqref{eq:23}, we have the expression of the Hessian, $\bm{H_i}$ for a specific $q_i$. 
\begin{equation}
\begin{split}
\bm{H_i} & = \nabla_{\bm{d}}^2 \left(\chi_i^2 \right)\\
 & = \frac{4}{I(q_i, \bm{d})J} \left[ \frac{I_{exp}(q_i)}{\sigma(q_i)} \right]^2 \Bigl\{ \log\left[ I(q_i, \bm{d}) \right] - \log\left[ I_{exp}(q_i) \right] \Bigr\} \sum_{k=1}^J \Bigl( \bm{a_k}\bm{a_k}^T + \bm{b_k}\bm{b_k}^T \Bigr)\\
 & + \frac{8}{\left[ I(q_i, \bm{d})J \right]^2} \left[ \frac{I_{exp}(q_i)}{\sigma(q_i)} \right]^2 \Bigl\{ 1 - \log\left[ I(q_i, \bm{d}) \right] + \log\left[ I_{exp}(q_i) \right] \Bigr\}\\
 & \times \sum_{k=1}^J \sum_{l=1}^J \Bigl[ \left(\bm{v}^T\bm{a_k} \right)\left(\bm{v}^T\bm{a_l} \right)\bm{a_k}\bm{a_l}^T + \left(\bm{v}^T\bm{a_k} \right)\left(\bm{v}^T\bm{b_l} \right)\bm{a_k}\bm{b_l}^T \\
 & + \left(\bm{v}^T\bm{b_k} \right)\left(\bm{v}^T\bm{a_l} \right)\bm{b_k}\bm{a_l}^T + \left(\bm{v}^T\bm{b_k} \right)\left(\bm{v}^T\bm{b_l} \right)\bm{b_k}\bm{b_l}^T \Bigr]
\end{split}
\label{eq:26}
\end{equation}
Due to linearity, the overall Hessian, $\bm{H}$, can be written as the sum of all the $\bm{H_i}$'s. 
\begin{equation}
\bm{H} = \nabla_{\bm{d}}^2\left(\chi^2 \right) = \frac{1}{n-1}\sum_{i=1}^n \nabla_{\bm{d}}^2 \left(\chi_i^2 \right) = \frac{1}{n-1}\sum_{i=1}^n \bm{H_i}
\label{eq:27}
\end{equation}
\begin{flushright}
$\blacksquare$\\
\end{flushright}

For typical numerical consideration, $\bm{H} \in \mathbb{R}^{N\times N}$ where $N \approx 10^5$, $J \approx 10^3$ and $n \approx 10^2$ at one \textit{FM} step. The computational complexity of $\bm{H}$ is $\mathcal{O}(nN^3J^2)$ which is very expensive and consumes memory beyond typical RAM in the computing cluster. \\

\subsection{The Customized $\chi^2$ Distribution}

The widely-used linear $\chi^2$ metric used for SAXS data is defined as follows. 
\begin{equation}
\chi_{Lin}^2 = \frac{1}{n-1} \sum_{i=1}^n \left[ \frac{I(q_i, \bm{d}) - I_{exp}(q_i)}{\sigma(q_i)} \right]^2
\end{equation}
Statistically, $(n-1)\chi_{Lin}^2$ follows the $\chi^2$-distribution with $n-1$ degrees of freedom. 
\begin{equation}
(n-1)\chi_{Lin}^2 \sim \chi_{n-1}^2 \sim \Gamma\left(\frac{n-1}{2}, \frac{1}{2}\right)
\end{equation}
where $\Gamma(\alpha, \beta)$ is the Gamma distribution with parameters $\alpha$ and $\beta$. The individual $\chi_{Lin, i}^2$ follows the $\chi^2$-distribution with 1 degree of freedom. 
\begin{equation}
\chi_{Lin, i}^2 = \left[ \frac{I(q_i, \bm{d}) - I_{exp}(q_i)}{\sigma(q_i)} \right]^2 \sim \chi_1^2 \sim \Gamma\left(\frac{1}{2}, \frac{1}{2}\right),
\end{equation}

We would like to derive the distribution of our customized $\chi^2$ metric defined in Eq. \eqref{eq:5}. For an experimental measurement at $q_i$, the real scattering amplitude $X = I(q_i)$ is a random variable following the normal distribution with $(\mu, \sigma) = (I_{exp}(q_i), \sigma(q_i))$. We now seek the distribution of a new random variable $Y$ defined as 
\begin{equation}
Y = \left(\frac{\log_{10}X - \log_{10}\mu}{\sigma'}\right)^2 = \left(\frac{\mu}{\sigma}\right)^2\Bigl(\log X - \log\mu \Bigr)^2
\end{equation}
where $\sigma'$ is defined in Eq \eqref{eq:2}. From the distribution of $X$, we have 
\begin{equation}
\int_{-\infty}^{\infty} \frac{1}{\sqrt{2\pi\sigma^2}}\exp\left[-\frac{1}{2\sigma^2}(x - \mu)^2 \right]dx = 1
\end{equation}
Given that the experimental S/N ratio is high and the measurement $\mu$ is a positive large number, we apply integral transformations: 
\begin{equation}
1 = \int_1^{\infty} \frac{2}{\sqrt{2\pi\sigma^2}}\exp\left\{-\frac{\mu^2}{2\sigma^2}\left[e^{(\log x - \log\mu)} - 1\right]^2 \right\}dx
\label{eq:33}
\end{equation}
Let $t  = \log x$ and $t' = t - \log \mu = \sigma\sqrt{y} / \mu$. Eq. \eqref{eq:33} becomes
\begin{equation}
\begin{split}
1 & = \int_{-\log\mu}^{\infty} \frac{2}{\sqrt{2\pi\sigma^2}}\exp\left[-\frac{\mu^2}{2\sigma^2}\left(e^{t'} - 1\right)^2 \right]\mu e^{t'} dt'\\
 & = \int_0^{\infty} \frac{1}{\sqrt{2\pi}}\exp\left[-\frac{\mu^2}{2\sigma^2}\left(e^{\frac{\sigma}{\mu}\sqrt{y}}-1 \right)^2 \right]e^{\frac{\sigma}{\mu}\sqrt{y}}\frac{1}{\sqrt{y}}dy
\end{split}
\label{eq:34}
\end{equation}
We define the signal-to-noise ratio (S/N) as $r = \mu / \sigma$ and re-write Eq. \eqref{eq:34} as 
\begin{equation}
\int_0^{\infty} \frac{1}{\sqrt{2\pi y}} \exp\left[ -\frac{1}{2}r^2\left(e^{\sqrt{y}/r}-1\right)^2 + \frac{\sqrt{y}}{r}\right]dy = \int_0^{\infty} f(y)dy = 1
\end{equation}
Therefore, the probability distribution function (p.d.f.) of the variable $Y$ is the following. 
\begin{equation}
f(y, r) = \frac{1}{\sqrt{2\pi y}} \exp\left[ -\frac{1}{2}r^2\left(e^{\sqrt{y}/r}-1\right)^2 + \frac{\sqrt{y}}{r}\right], \quad y, r > 0
\label{eq:36}
\end{equation}
Notice that as $y \to \infty$, the exponent approaches $-\infty$ and the function goes to zero double exponentially. 
\begin{equation}
\lim_{y\to\infty}f(y, r) \approx \lim_{y\to\infty}\frac{1}{\sqrt{y}}e^{-e^{2\sqrt{y}}} \to 0
\end{equation}
This behavior is similar to the $\chi_1^2$ distribution except that $\chi_1^2$ approaches $0$ as $x\to\infty$ only exponentially. On the other hand, as $y\to 0$, we can Taylor expand Eq. \eqref{eq:36} to $\mathcal{O}(y)$. The exponent $e^{[\hdots]}$ in Eq. \eqref{eq:36} is 
\begin{equation}
[\hdots] = -\frac{1}{2}r^2\left(e^{\sqrt{y}/r}-1\right)^2 + \frac{\sqrt{y}}{r} \approx -\frac{1}{2}y + \frac{\sqrt{y}}{r} 
\end{equation}
The term $\frac{\sqrt{y}}{r}$ is equal to $\log x - \log\mu$ according to our substitution, is relatively insensitive given large signal-to-noise ratio $r$ and can be viewed as a slow-varying constant compared to $\frac{1}{2}y$. Therefore, as $y\to 0$, we have 
\begin{equation}
\lim_{y\to 0}f(y, r) \sim \frac{1}{\sqrt{2\pi}}y^{-\frac{1}{2}} e^{-\frac{1}{2}y}
\end{equation}

\begin{table}
	\caption{The first and second moment of the distribution $f(y, r)$ in Eq. \eqref{eq:36} at a few signal-to-noise ratios, $r$.}
	\centering
	\begin{tabular}{ccccccccc}
		\toprule
		 & \multicolumn{1}{c}{$\chi_1^2$} & \multicolumn{7}{c}{$f(x, r)$}\\
		 \cmidrule{3-9}
		 & & $r=0.5$ & $r=1$ & $r=5$ & $r=10$ & $r=50$ & $r=150$ & $r=500$ \\
		\midrule
		First Moment & $1.0$ & $0.233$ & $0.387$ & $0.761$ & $0.864$ & $0.969$ & $0.989$ & $0.997$\\
		Second Moment & $3.0$ & $0.096$ & $0.300$ & $1.488$ & $2.046$ & $2.760$ & $2.917$ & $2.975$\\
		\bottomrule
	\end{tabular}
	\label{tab:1}
\end{table}

\begin{figure}[H]
\centering
\includegraphics[width=\textwidth]{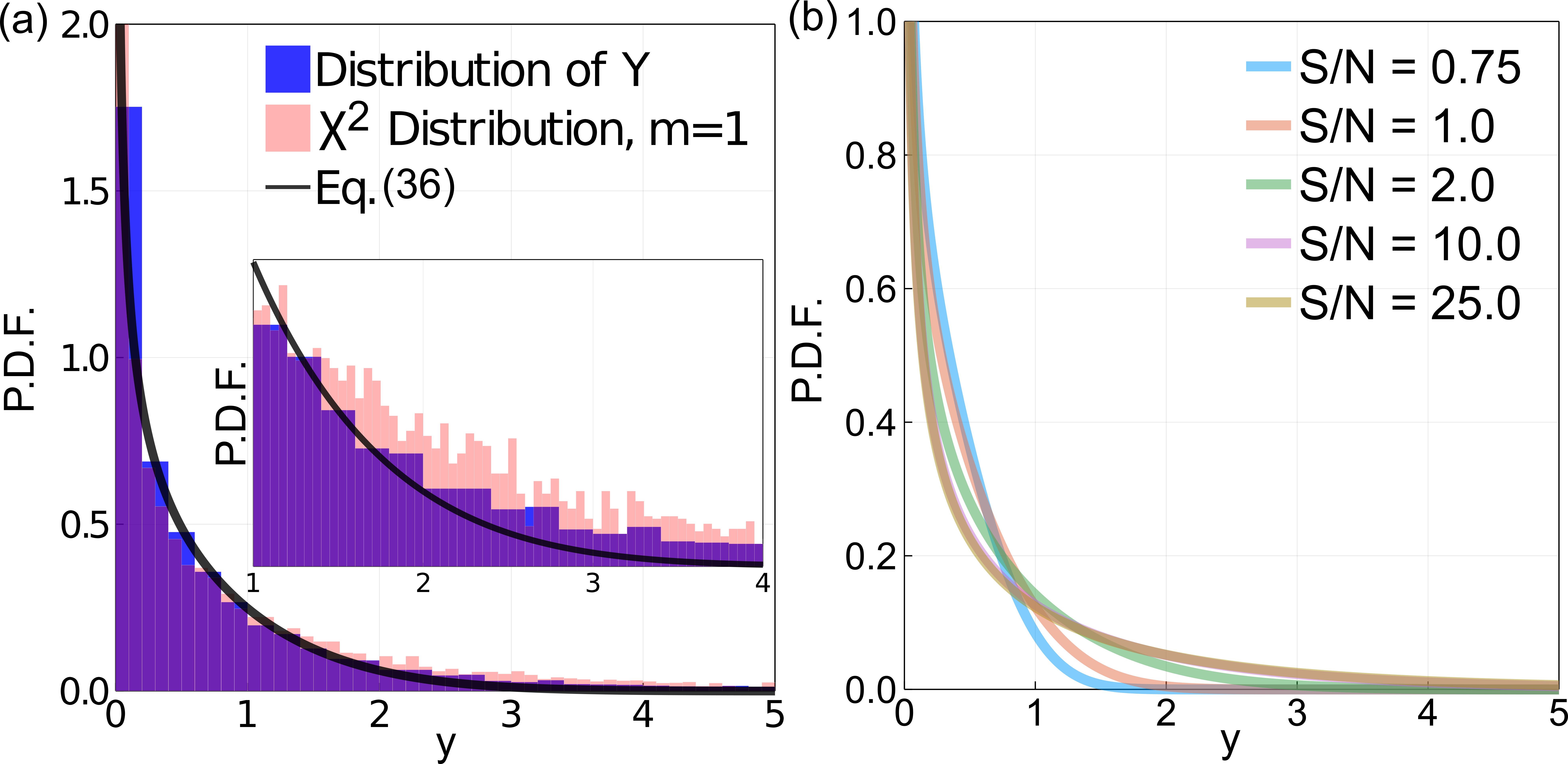}
\caption{(a) We simulate a normally distributed random variable $X$ and convert it to the linear and customized $\chi^2$ metric shown in pink and blue respectively, using $\mu=3.39\times 10^5$ and $\sigma=1.67\times 10^5$ from one experimental point at $q=1.25\angstrom^{-1}$. The solid line is the distribution of Eq. \eqref{eq:36} with $r = \mu/\sigma = 2.03$. The inset focuses on $y > 1$ and shows the differences between two distributions. (b) The distribution of Eq. \eqref{eq:36} with different signal-to-noise values ranging from $0.75$ to $25.0$. }
\label{fig:1}
\end{figure}

The distribution of random variable $Y$, $f(y, r)$, is similar to $\chi^2$-distribution with $1$ degree of freedom in both limits of $y$. The left panel of Fig. \ref{fig:1} shows the simulated distributions of the linear and customized $\chi^2$ metrics, corresponding to $X$ and $Y$ respectively using the empirical data at $q = 1.25\angstrom^{-1}$. The solid line was computed using Eq. \eqref{eq:36} showing good agreement with the distribution of $Y$, our customized $\chi^2$ metric. The right panel of Fig. \ref{fig:1} shows five distributions from Eq. \eqref{eq:36} with different signal-to-noise ratios, $r$. The effect is more significant only when $r \leq 2.0$, which is below typical experimental measurements in the wide-angle regime. The first and second moments of $f(y, r)$ are reported in Table \ref{tab:1} with various signal-to-noise ratios. As $r$ increases, the distribution $f(y, r)$ approaches $\chi^2$-distribution with one degree of freedom.
\begin{equation}
\lim_{r\to\infty} f(y, r) = \frac{1}{\sqrt{2\pi}}y^{-\frac{1}{2}} e^{-\frac{1}{2}y}
\label{eq:40}
\end{equation}
Notice that $f(y, r)$ depends on $r$, so the optimal goodness-of-fit statistics might not necessarily be $1.0$ as in the linear $\chi^2$ case. As a result, this metric in Eq. \eqref{eq:1} gives reasonably similar $\chi^2$-distribution statistics while avoiding biases at low-$q$ profiles and improves numerical stability. 
\begin{flushright}
$\blacksquare$
\end{flushright}

\subsection{Solution Space}

We now focus on how the 3D solution space $\mathcal{S}$ is transformed with the starting molecular shape, $\mathcal{P}$. We define $\mathcal{E}$ as the ensemble space or the conformational space of the molecule and $\mathbb{R}$ as the real space. The goal is now to find the cardinality of the solution space with and without the shape $\mathcal{P}$. If there is no experimental measurement and molecular shape, the  solution space can be written as follows. 
\begin{equation}
\mathcal{S} = \mathcal{E} \otimes \mathbb{R}^3 \otimes [0, 2\pi]\otimes [0, \pi]\otimes \{+1, -1 \}
\end{equation}
The $\mathbb{R}^3$ is the translation symmetry in 3D while $[0, 2\pi]$ and $[0, \pi]$ denote the rotational symmetry. The term $\{+1, -1\}$ is the chiral symmetry. Therefore,
\begin{equation}
\text{card}(\mathcal{S}) = 2\text{card}(\mathcal{E})\text{card}(\mathbb{R})^3\text{card}([0, 2\pi])\text{card}([0, \pi])
\end{equation}
Using the bijective mapping function of $f(x) = \tan\left(x-\frac{\pi}{2} \right)$,  $\text{card}([0, \pi]) = \text{card}(\mathbb{R})$. With the same reasoning and bijective function $f(x) = \tan\left[ \frac{1}{2}\left(x-\frac{\pi}{2} \right)\right]$, we have $\text{card}([0, 2\pi]) = \text{card}(\mathbb{R})$ which is uncountably infinite. Therefore, 
\begin{equation}
\text{card}(\mathcal{E}) = 2\text{card}(\mathcal{E})\text{card}(\mathbb{R})^5
\end{equation}
Given the molecular shape $\mathcal{P}$, the solution space is confined with known translational and rotational symmetry. The cardinality of the conditional solution space is then
\begin{equation}
\text{card}(\mathcal{S}|\mathcal{P}) = 2\text{card}(\mathcal{E}|\mathcal{P}) \text{card}(\mathbb{R}|\mathcal{P})^5
\end{equation}
Assume that $\mathcal{P}$ is not a perfect sphere without any 3D orientational information, i.e. $\mathcal{P} \not\in \{\mathcal{B}(r): \forall r >0\}$ where $\mathcal{B}(r)$ is the 3D sphere with radius $r$. Since we know the translation and orientation, given a non-sphercal prior shape $\mathcal{P}$, $\text{card}(\mathbb{R}|\mathcal{P}) \approx 1$. 
\begin{equation}
\text{card}(\mathcal{S}|\mathcal{P}) \approx 2\text{card}(\mathcal{E}|\mathcal{P})
\label{eq:45}
\end{equation}
Eq. \eqref{eq:45} suggests that solution space is greatly reduced to the $\text{card}(\mathcal{E}|\mathcal{P})$ up to a factor of $2$ due to chiral symmetry. Furthermore, $\text{card}(\mathcal{E}|\mathcal{P}) = \text{card}(\mathcal{E})$ because we don't know anything about the ensemble space as long as the molecular shape $\mathcal{P}$ contains or shrouds the intrinsic 3D ensemble space. This is usually true since $\mathcal{P}$ is derived from SAXS profile, representing the molecular shape.
\begin{equation}
\text{card}(\mathcal{S}|\mathcal{P}) \approx 2\text{card}(\mathcal{E}) \quad \forall \mathcal{P} \not\in \{\mathcal{B}(r): \forall r >0\}
\end{equation}
As a result, with a high-confident molecular shape/envelope $\mathcal{P}$, \textit{FM} looks for one solution within the conformational space $\mathcal{E}$ up to the chirality.

\section{Frequency Marching Algorithm}

See Algorithm \ref{alg:1} for the pseudo-code of the proposed \textit{Frequency Marching} (\textit{FM}) algorithm. It takes the 3-column experimental SWAXS profile and 3D electron density (denoted as "mrc") as two main inputs. It requires the starting number of voxels and $q_{max}^{(0)}$ for the zeroth \textit{FM} step. The zoom ($z$) parameter determines how much further the algorithm marches to at certain \textit{FM} step $i$. Moreover, the scheduling of the following parameters is necessary because as \textit{FM} marches into higher angle, accuracy is undermined by the use of the same set of parameters under the orientational average scheme. Different parameters should be used to bridge between the previous \textit{FM} step and the next. The higher resolution, the weaker smearing power should be applied. We report the effects of zoom, smearing power and width of solvent layers on the SWAXS profile in Fig. S1-S3. The electron density of the solvent background on a voxel should also be scaled according to the voxel size, with a default value of $0.335 e/\angstrom^3$. Therefore, heuristically and from supporting figures, we used the following scheduling strategies for the number of orientations ($J_i$), smearing power ($\sigma_i$), maximum iterations ($m_i$) and maximum electron density on a voxel ($ub_i$) at \textit{FM} step $i$. 
\begin{equation}
m_i = m_0 e^{i^3/1000}
\label{eq:47}
\end{equation}
\begin{equation}
\sigma_i = a + b e^{-\mu(i-1)},
\end{equation}
where $0.3 \leq a+b \leq 1.0$ and $a, b \geq 0$.
\begin{equation}
ub_i = ub_0 \times dxdydz
\end{equation}
where $dxdydz$ is the voxel size at step $i$ and in normal case the voxel is a cube, $dx=dy=dz$. $ub_0$ is the maximum allowed average electron density in $e/\angstrom^3$. 
\begin{equation}
J_i = \max\left(J_0, J_0 q_{max, i}^{5/2} \right)
\end{equation}
$J_i$ should be proportional to $(qD)^2$ where $D$ is the maximum molecular dimension, which we dropped here and use $q^{5/2}$ to adapt for good precision at wide-angle regime. We apply \textit{Adam} algorithm in the \textit{carving} step\cite{Kingma2014}.

\begin{algorithm}
\caption{Frequency Marching (FM) Algorithm}
\begin{algorithmic}[1]
\Require mrc from SAXS, SWAXS profile, fmsteps, zoom ($z$)
\Require Starting number of voxels, $q_{min}$, $q_{max}^{(0)}$
\Require Schedule the following: $J_i$, $\sigma_i$, maximum iterations ($m_i$), upper bound($ub_i$)\\
Initialize parameters and scheduling\\
Calculating the SWAXS profile from mrc\\
Match the number of electrons in mrc
\For{i = 1 : fmsteps}
	\State $q_{max}^{(i)} \leftarrow zq_{max}^{(i-1)}$
	\State Smear mrc w.r.t. $\sigma_i$
	\State Split mrc w.r.t. $z$
	\State \textit{Carve} mrc w.r.t. $J_i$ and $m_i$
\EndFor\\
\Return mrc, statistics, SWAXS profile
\end{algorithmic}
\label{alg:1}
\end{algorithm}

\begin{figure}[h]
\centering
\includegraphics[width=\textwidth]{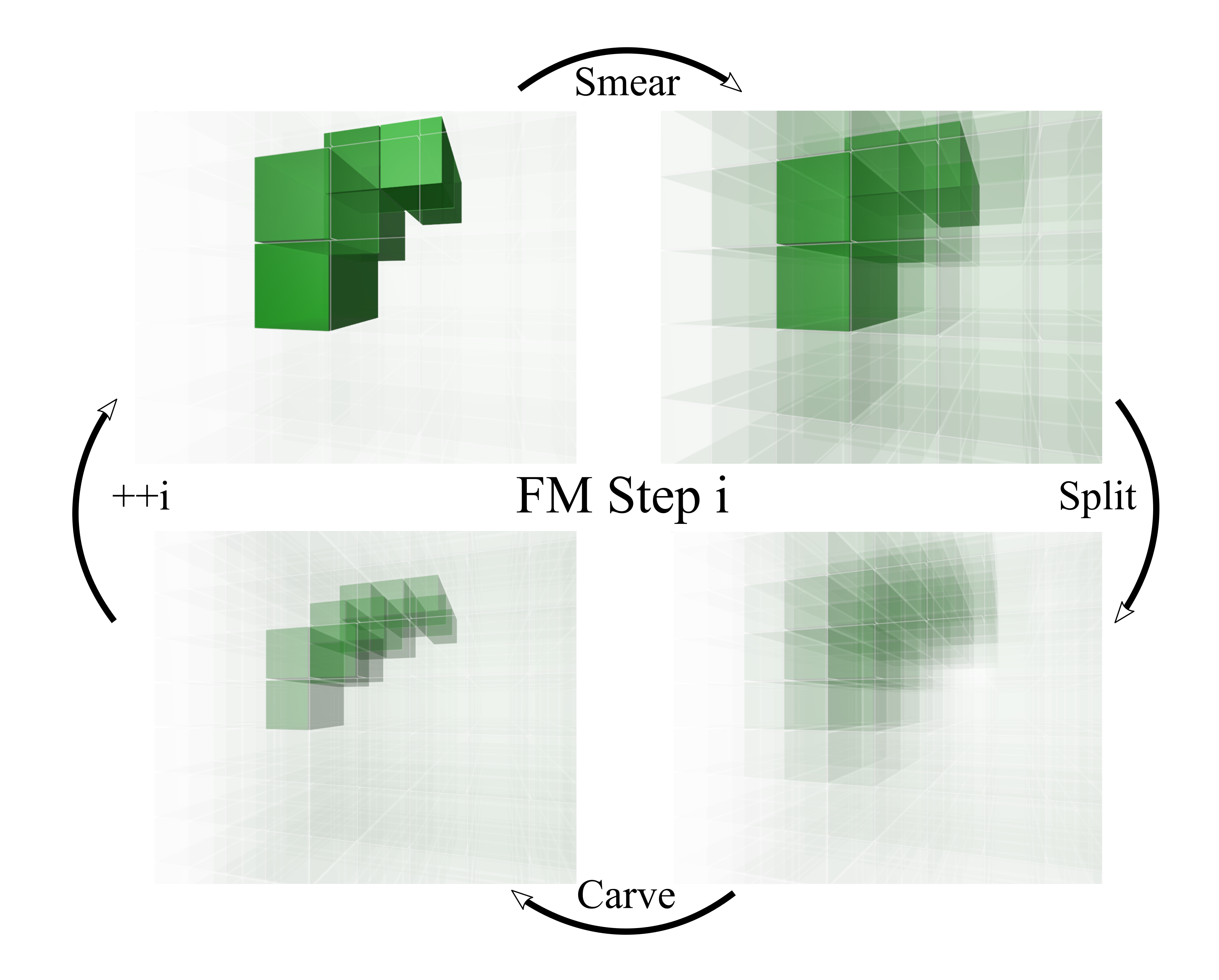}
\caption{The schematics of Algorithm \ref{alg:1}. At each \textit{FM} step, the electron density is smeared before being split. After splitting, the algorithm \textit{carves} the electron density to reveal finer structures on length scales defined by $q_{max}^{(i)}$ and proceeds to the next \textit{FM} step.}
\label{fig:bld}
\end{figure}

Fig.~\ref{fig:bld} shows the illustration of Algorithm \ref{alg:1} line 6-8. There are two levels in the \textit{FM} concept. The first is to \textit{carve} out the 3D electron density at $i$\textsuperscript{th} \textit{FM} step for a given $q_{max}^{(i)}$. This process becomes more computationally expensive as higher $q$ is achieved. Therefore, in each \textit{FM} step, it is optimal to start from the molecular envelope as close to the solution as possible, i.e. low starting $\chi^2$. It is possible to start with the lowest resolution envelope and march one step to $q_{max}$. However, because of the large number of voxels, optimal $\chi^2\approx 1$ is not achieved within a reasonable number of iterations. This is the reason for the second level: \textit{marching}, which quickly generates a closer starting envelope for the next \textit{FM} step to reduce the computation time.

\section{Material and Methods}

\subsection{Sample Preparation}

In this work, we used 5 nucleic acid constructs: 25-base-paired double-stranded DNA (25bp dsDNA), RNA hairpin with 12bp double-stranded (hpRNA) region, triple-stranded RNA (tsRNA), metastasis-associated lung adenocarcinoma transcript 1 (MALAT1) and single-stranded RNA (ssRNA) with 30 mixed nucleobases. The 25bp dsDNA contains one strand with sequence: 5'-GCA TCT GGG CTA TAA AAG GGC GTC G-3' along with its complement. The two strands were purchased from IDT (Coralville, IA), annealed, purified and then buffer exchanged to a solution containing 100 mM NaCl, 10 mM sodium 3-(N-morpholino)propanesulfonic acid (Na-MOPS) and 20 $\mu$M EDTA, pH 7.0. The RNA hairpin sequence is 5’-GGG CUA GAG AGA GAA AGU UUC GAC UUU CUC UCU CUA GCC C-3’ and was synthesized by T7 transcription (Promega, Fitchburg, WI) and purified by Mono Q 5/50 GL anion exchange column (GE Healthcare, Chicago, IL). The RNA triplex was made by annealing the RNA hairpin with a third strand of triplex forming oligo (TFO), anti-sense 12-bp single-stranded RNA with corresponding sequence: 5’-UCU CUC UCU UUC-3’ \cite{Zhou2013, Devi2015}. The RNA triplex was further separated by the same anion exchange column. 
The single stranded RNA contains $30$ nucleotides with mixed sequence: 5'-AAG AAU AAA AGA GAA GCC ACC CCA CCC AGA-3' of no known secondary structure and was also purchased from IDT (Coralville, IA). The MALAT1 sample was prepared as described in\cite{Brown2014}. The final sample concentrations ranged from $100$ to $150$ $\mu$M to achieve adequate signal-to-noise ratio in the wide-angle regime.

\subsection{Solution X-ray Scattering Measurements}

The solution X-ray scattering data for all of our nucleic acid systems were acquired at 16-ID (LiX) beamline of National Synchrotron Light Source II (NSLS II)\cite{Yang2020} at Brookhaven National Laboratory. 60 $\mu$L of solution samples as well as their matching buffers were manually loaded onto the sample holder and measured in continuous-flow mode with five 1-second exposures. Data spanning q values of $0.005$ to $3.2 \angstrom^{-1}$ were recorded using a Pilatus3 1M (SAXS) and two Pilatus3 300K (WAXS) detectors (Dectris, Switzerland) in vacuum. The signal-to-noise ratio at $q = 1.5 \angstrom^{-1}$ was above $10$ for all of our sample conditions. On-site, data were quality controlled using the \textit{py4xs} Python package. The small-angle part of the profiles, spanning $q=0.005$ to $0.25 \angstrom^{-1}$ was used for the \textit{ab initio} electron density reconstruction, created using \textit{DENSS}\cite{Grant2018}. We generated $25$ reconstructions in parallel and averaged them as the starting molecular envelope for our \textit{FM} algorithm.

\section{Results}

\subsection{Frequency Marching Steps}

\begin{figure}[H]
\centering
\includegraphics[width=\textwidth]{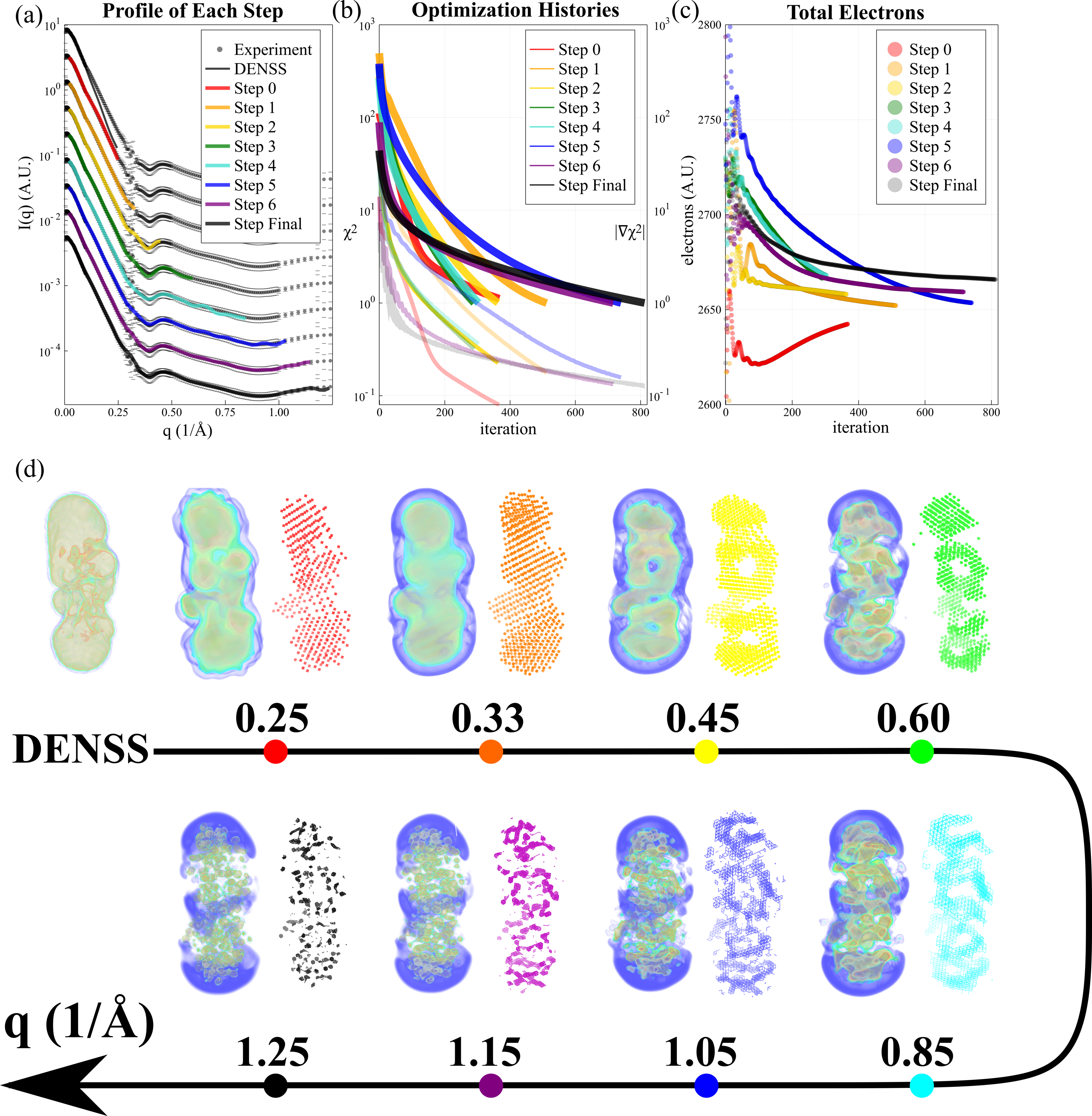}
\caption{(a) The fits from different \textit{FM} steps to the experimental SWAXS profile of 25bp dsDNA. The profiles are offset by different scale factors for easy visualization. The scattering intensity is shown using a logarithmic scale that magnifies subtle disagreements. (b) The refinement traces of $\chi^2$ (solid) and $|\nabla_{\bm{d}}(\chi^2)|$ (transparent) for all the \textit{FM} steps. Good convergence can be observed by decaying $\chi^2$ and L2-norm of the gradient within a few hundred iterations. (c) The total electrons for all \textit{FM} steps. (d) The density maps of the refined electron densities from different \textit{FM} steps. The density maps are presented with the same coloring scheme as the curves. The maps were generated using PyMOL (Schrodinger, New York, NY). Each subsequent density map varies in resolution and recapitulates the experimental data to a specified $q_{max}$. We construct a wire plot associated with each density map by extracting coordinates of the voxels that contain more than $80\%$ of the maximum density. These maps are shown with the same color code used in the top-left panel. The first density map is from \textit{DENSS} using small-angle data. See main text for detail. }
\label{fig:2}
\end{figure}

We first use the full small and wide angle (SWAXS) profile of 25bp dsDNA as a demonstration of concept. Fig. \ref{fig:2}(a) shows the progression of the algorithm, as fits to the experimental data from all the \textit{FM} steps. Different scaling constants are used in the figure for ease of visualization. All profiles are plotted as $\log_{10}(I)$ versus $q$ to highlight the subtle differences in the wide-angle scattering. Notice that the SAXS profile from \textit{DENSS} was computed using the orientational average formulation of Debye formula with buffer subtraction instead of Fast Fourier Transform (FFT)\cite{Grant2018}. Each voxel was used as the fundamental unit to build up the scattering profile. The zeroth step involves a quick refinement within the envelope using only SAXS data up to $q=0.25\angstrom^{-1}$. The resultant density serves as the starting point for the first \textit{FM} step and so on. For the convergence criteria of each \textit{FM} step, in addition to the maximum number of iterations, $m_i$ defined in Eq. \eqref{eq:47}, we used the first moment of the customized $\chi^2$ distribution in Eq. \eqref{eq:36} appropriate for the q-averaged signal-to-noise ratio from all the detected angles. We also require the L2-norm of the gradient in Eq. \eqref{eq:21} to be above a threshold of $0.05$ for computational efficiency since small gradient only improves the result marginally. Fig. \ref{fig:2}(b) shows the traces of $\chi^2$ (solid lines) and $|\nabla_{\bm{d}}(\chi^2)|$ (transparent lines) of all the \textit{FM} steps. For each \textit{FM} step, the convergence is demonstrated by the analytical gradient, which asymptotes within a few hundred iterations. The value of $\chi^2$ approaches its optimal value, $\chi^2\approx 1$ given the large signal-to-noise ratio of the data. The starting $\chi^2$ values are below $1000$. In the context of values selected by other structure determining algorithms at the start of iteration, this is a reasonable choice that demonstrates the  success of the starting molecular envelope derived from \textit{DENSS}. The large starting $\chi^2$ at subsequent \textit{FM} steps results from the inclusion of wider-angle data and the splitting of the 3D voxels. Fig. \ref{fig:2}(c) shows the traces corresponding to the total number of electrons for all the \textit{FM} steps. These sets of curves demonstrate that the results of the \textit{FM} algorithm suggest reasonable numbers of electrons to match the experimental $I(0)$ but also demonstrates their rearrangements to reflect the subtle features in the wide-angle regime. This \textit{carving} process can be visualized in the short movies in the supporting material. 

Fig. \ref{fig:2}(d) shows the refined 3D electron density derived from \textit{FM} corresponding to consideration of data with different/increasing $q_{max}^{(i)}$. For clarity, each step is designated with the same color code used in Panel (a). The first density map corresponds to the \textit{DENSS} reconstructed molecular envelope. For each subsequent \textit{FM} step, one refined electron density map is shown (left most curve of each pair). The computed SWAXS profile from these structures is shown as the solid lines in the top-left panel. We use the contour coloring scheme in PyMOL (Schrodinger, New York, NY) for the presentation. The wire plots on the right, with colors corresponding to the \textit{FM} step number, show the positions of those voxels containing more than $80\%$ of the maximum density. This visualization highlights the electron-dense structures within the molecular envelope that contribute the most to the SWAXS features. For the zeroth and first steps, the algorithm only "marches" to $q=0.33\angstrom^{-1}$, which is still in the small-angle regime; because of the correspondingly low resolution, these steps simply refine the low-resolution density. At higher values ($q=0.45\angstrom^{-1}$) the scattering profile of the 25bp dsDNA profile exhibits features. Here we begin to see a hole and the emergence of some periodic structures. These features become \textit{carved} out of the lower density representation. The electron-dense region might represent the phosphate-rich (electron dense) backbone or the cations attracted to the negatively charged DNA surface. We cannot distinguish them using this implicit electron modeling scheme; density is reflected, but the identity of the underlying atom is not. It is also worth noting that the electron density near the molecular boundary approaches zero and even turns slightly negative relative to the bulk solvent. This molecular shell was rapidly \textit{carved} away as a result of the large density gradient near the boundary when higher-angle data (contrast between $I(0)$ and high-$q$ profile) are included. Previous modeling methods, performed with atomically detailed models, also require that such a layer exists in order to match the experimental data\cite{Chen2014, Knight2015}. As \textit{FM} marches to $q=1.05\angstrom^{-1}$ and beyond, the overall density becomes fractionated while still maintaining the overall structure from previous \textit{FM} steps. We did not require connectedness of the electron density because it undermines the 3D gradient profile and convergence as resolution increases. Finally, the shallow or less-dense region in the middle of the structure corresponds to the dsDNA base pairs, which are hydrophobic and less electron dense than the backbone and bulk buffer. As a result, given a SAXS envelope, we demonstrated the concept and feasibility of the \textit{FM} algorithm in refining 3D electron density against the full spectrum of solution X-ray scattering profile.

\subsection{Structured, Unstructured and Biological RNA}

\begin{figure}[H]
\centering
\includegraphics[width=\textwidth]{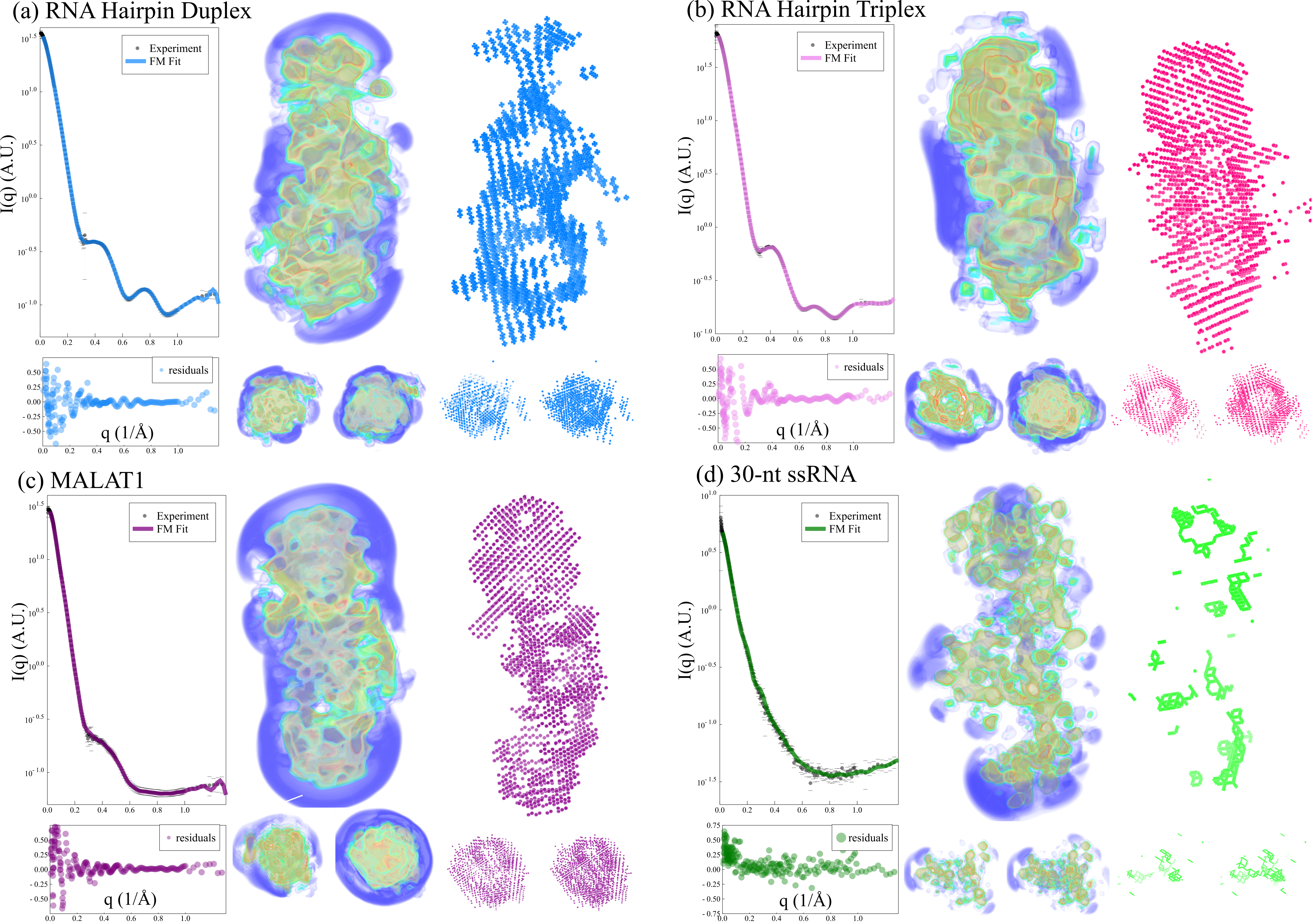}
\caption{The application of \textit{FM} algorithm to different systems: (a) RNA hairpin (hpRNA), (b) RNA triplex (tsRNA), (c) metastasis-associated lung adenocarcinoma transcript 1 (MALAT1) and (d) a short strand of single-stranded RNA (ssRNA). Each panel displays the experimental data and \textit{FM} fit, with residuals computed from Eq. \eqref{eq:51}, the \textit{FM} refined electron density and corresponding wire plot as in Fig. \ref{fig:2}. The two cross-sections associated with either density map or wire plot represent sections through the midsection (left) or across the top (right) viewing into the central axis of the molecule. See main text for detailed explanation.}
\label{fig:3}
\end{figure}

With its utility to a standard structure (B-form DNA) demonstrated above, we now apply \textit{FM} to a variety of structured, unstructured and biological RNA systems. The same procedure was followed to obtain a low resolution molecular envelope using the experimental small-angle data up to $q=0.25\angstrom^{-1}$. Seven \textit{FM} steps were taken to refine the 3D electron density, out to a maximum value of $q=1.25\angstrom^{-1}$. Fig. \ref{fig:3} shows the superposition of the fit of the final refinement step of the \textit{FM} fit with the raw experimental data and the corresponding real space densities for (a) an RNA hairpin (hpRNA), (b)an RNA triplex (tsRNA), (c) an important regulatory element: metastasis-associated lung adenocarcinoma transcript 1 (MALAT1) and (d) a single strand of RNA (ssRNA). In each panel, the raw experimental SWAXS profile and the  \textit{FM} fit are displayed on a logarithmic scale. The residuals are computed and plotted below using 
\begin{equation}
\text{residual}(q_i) = \frac{\log_{10}\left[ I(q_i, \bm{d}) \right] - \log_{10} \left[ I_{exp}(q_i) \right]}{\sigma'(q_i)}
\label{eq:51}
\end{equation}
with the same notation as in Eq. \eqref{eq:5}. The density map and the wire plot showing the main structures are also presented, with molecular cross-sections in the middle (left) and at the top (right) (a view along the central axis). Each structure presents a distinct WAXS profile, and we now compare them one by one.

We start with the hpRNA, which contains a 18bp duplex stem and a loop formed by 4 nucleobases. Because of the length of the stem helix, its fine periodic features dominate the scattering profile. Because of their physical dimensions, they are mapped to the wide-angle regime (features of the canonical A-form RNA duplex, such as radius, groove width etc. occur on length scales at or below 10 \angstrom). The refined density shown in the top-left panel of Fig. \ref{fig:3} recapitulates the loop at the top as well as 3D electron-dense regions that correspond to the backbone or ion cloud in the stem. Holes or empty regions within the structure suggest a relatively sparse density along the central axis of the molecule, as observed for the 25bp dsDNA duplex discussed in the last section.

Moving to the second structure, the hairpin triplex, we rely on intuition gained from previous work using machine learning methods to interpret the WAXS features. For an RNA duplex \cite{Chen2020}, we 'learned' that the second peak at $q\approx 0.7\angstrom^{-1}$ reflects the major groove. It is known that the third RNA strand involved in creating a triplex (three stranded structure) from a duplex (two stranded structure), the so called triplex forming oligo or TFO, occupies the RNA major groove via stabilization from consecutive base triples. Thus we expect a modification in this feature of the scattering profile. The SWAXS profile of our tsRNA  construct is shown in the top-right panel of Fig. \ref{fig:3}, and displays the expected and significant difference in the wide-angle feature. In this case, the tsRNA is made from hpRNA by locking 12 out of 18 base pairs in the stem right below the loop (the TFO is 12 bases long). Therefore, the SWAXS profile reflects $67\%$ RNA triplex and $33\%$ RNA duplex structures. Despite this nonuniformity of structure, a clear change is seen near $q\approx 0.7\angstrom^{-1}$, resulting from    disruption of RNA major groove periodicity. Altogether, the \textit{FM} refined electron density agrees with the known design of our tsRNA. The more electron-dense structures near the loop region (top) of the molecule reflect the triplex, while the lower part of the stem, lacking the TFO, is less electron dense. Interestingly, the cross-sections shown below the panel have a "vacant tube" along the central axis, implying a large electron density difference (higher contrast) between the phosphate backbone structures and the base triples/pairs in the center. This density vacancy should appear in the RNA duplex due to the same contrast and reasoning (panel a); yet it does not appear in the \textit{FM} refined density. We speculate that this difference involves the accessibility of the central region through the RNA major groove. \textit{In vitro}, this region is occupied by water or cations that are attracted to the negative RNA surface, especially the highly negative major groove. In contrast, in the tsRNA, the major groove is occupied by a third strand TFO, blocking all the ion and water access into the "tube" region, leaving a sparser electron density.

We now discuss the MALAT1 molecule, which is a self-forming bipartite RNA triplex present in many long non-coding RNAs (lncRNAs). This important, biologically relevant motif inhibits nuclear RNA decay\cite{Brown2014}. The SWAXS profile and \textit{FM} refined density are shown in Fig. \ref{fig:3}(c). The experimental SWAXS profile lacks the major groove feature due to triplex formation in the lower part of the molecule. The \textit{FM} refinement shows the denser helical structures in the lower 2/3 of the structure, with the third strand in the middle part of the density map. Our \textit{FM} reconstruction of  MALAT1 has an intrinsic bend, in agreement with the crystal structure (PDB:4PLX). 

The above described comparisons among hpRNA, tsRNA and MALAT1 demonstrate the power of the WAXS experiments coupled with the \textit{FM} algorithm. Although these three molecules have similar overall cylindrical shapes, they differ substantially in their finer structural detail, which, in turn is reflected in their very different biological roles. This finer scale information is invisible to SAXS experiments.

The three RNA systems discussed above are well-structured and relatively rigid. They are constrained by the formation of base pairs or base triples so the \textit{FM} refined electron densities can be well-described with dense and sparse regions. To fully explore the limits of the \textit{FM} approach, we now apply it to a less structured RNA system, ssRNA. In the absence of secondary structure, the ssRNA is highly flexible \textit{in vitro}. To avoid potential base-stacking interactions that might affect the path of the phosphate backbone, we used a mixed sequence construct. The SWAXS profile and the density map are shown in Fig. \ref{fig:3}(d). No apparent features are seen in the SWAXS curves, in contrast with the structured RNA molecules, suggesting the absence of periodicity in the ssRNA due to its flexibility. The \textit{FM} refined electron density shows mostly unstructured regions with limited connectedness. Despite the presence of a few, denser voxels, most of the voxels have comparable electron densities. The wire plot for the density map captures only the former voxels while the rest are filtered out because of their low and uniform densities. The cross sectional views are mostly featureless, providing additional support for the view that the flexibility of the ssRNA smears out any structural periodicity.

\section{Discussion}

\subsection{From SAXS to WAXS Applications}

Taken together, our results show that solution X-ray scattering into the wide-angle regime is a powerful sensor of structural periodicities on short (less than 10 \AA) length scales. This length scale is ideally matched to structural features of nucleic acid systems and can be readily interpreted using the \textit{FM} algorithm introduced here. Prior work, yielding quantitative interpretation of  WAXS profiles 
required incorporation of molecular dynamics (MD) simulation to either fit the measurements\cite{Chen2019}, select important features for machine learning algorithms \cite{Chen2020}. 
However, many biological systems and phenomena may prove challenging to model by MD simulations, due to long simulation times or artifacts from force fields. Such systems, which display periodicities on the salient length scales,  include transitions of protein secondary structures, formation of RNA triplexes from single-stranded RNA tail in pseudo-knots, riboswitches or other structuring, subtle conformational changes induced by bound ligands or ions. These systems are  pose fundamental biological problems, and their solution is required to improve our detailed understanding of the workings of functional molecular machines. If these conformational changes increase or disrupt structural periodicities, WAXS measurements along with our proposed \textit{FM} algorithm can interpret and distinguish the difference in conformations. Despite the inability to assign explicit atoms or residues to the refined electron density, this improved visualization has great potential in sensing fine electron density rearrangements, interior bending, change in secondary structures and so on for bio-molecules that resist crystallization or are too small to be visible in cryoEM.

\subsection{Resolution Limit}

Because solution X-ray scattering experiments represent orientational averages, they do not yield a well-defined resolution of the type quoted in X-ray crystallography or cryoEM experiments. Although a general formula $d = 2\pi/q_{max}$ exists, linking angle and real space distance $d$, the so-called resolution, what is really detected is the smallest structural periodicity reported by the solution X-ray scattering experiments. For very well-structured macromolecular systems, finer periodicity can be inferred. For example, in the RNA duplex system, which display order over a large length scale, averaged geometries of di-base-pair step, such as helical rise and twist can be inferred from the solution X-ray scattering in the wide-angle regime. 

In the \textit{FM} reconstructed density, it is impossible to assign atoms or residues because of the intrinsic transformation from pair-distances to profiles. However, in well-structured systems with significant features, the finer periodicity will appear in the \textit{FM} reconstruction. In theory, the \textit{FM} algorithm can be applied to any $q > 0$. However, as higher $q$ is reached, implicit density modeling becomes less precise. Beyond $q = 1.5\angstrom^{-1}$, complex and explicit ion-solute and water-solute interaction might be essential to model. Moreover, only at high-$q$ does one need to consider the "negative" electron density relative to the bulk solvent. Finally, \textit{FM} does not account for effect of experimental sample concentrations; the "effective" sample concentration using our implicit modeling is high. This might be the reason why the electron density becomes fractionated when wider-angle data is included in the \textit{carving} as background and contrast shift. As a result, the authors would impose a limit at $q=1.5\angstrom^{-1}$ as the "resolution" limit for state-of-the-art WAXS modeling using implicit densities.

\subsection{Limitations}

Despite the successful application of \textit{FM} algorithm to the structured nucleic acid systems, there are still limitations associated with either the \textit{FM} algorithm, macromolecular systems or the nature of solution X-ray scattering. First of all, although it's possible to estimate the electron densities on the voxels, the $q$-dependent atomic scattering form factors, representing the electron density distribution within the atom itself, are not accounted for in our implicit modeling scheme. This effect becomes more significant at wide angles, $q\approx 1.5 \angstrom^{-1}$ and beyond. Second, for flexible molecular systems, such as ssRNA, which exists as conformational ensembles with significantly diverse structures \textit{in vitro}, \textit{FM} will not work because the refined electron density reflects the ensemble averaged density within the given molecular envelope. There is no way to distinguish different conformations without more advanced atomic modeling. Third, it is essential that the starting shape $\mathcal{P}$ breaks rotational symmetry as discussed in Section 2.4. Finally, due to the nature of solution X-ray scattering, the reconstructions or refinements are not unique. The limitations of state-of-the-art reconstruction methods apply to \textit{FM}. However, given a starting structure that resembles the molecule with high-confidence, the refined electron density within will be consistent due to the analytical gradient and deterministic nature of \textit{FM} algorithm.



\section{Conclusion}

In this work, we present the \textit{Frequency Marching} algorithm that refines the 3D electron density distributions \textit{in vitro} against solution X-ray scattering profiles into the wide-angle regime ($q > 0.3\angstrom^{-1}$) with a customized but statistically similar $\chi^2$. This algorithm successfully recapitulates both structured and unstructured RNA system, with more information provided in the former case, demonstrating that finer structures can be refined from high quality SWAXS data.  We envision further applications of \textit{FM} to other macromolecular systems that display structural periodicities on biologically important, nearly molecular length scales.

\section{Acknowledgment}

This work was supported by NIH Grant R35GM122514. Support for work performed at the CBMS beam line LIX (16ID) at NSLS-II is provided by NIH - P30 GM133893, S10 OD012331 and BER- BO 070. NSLS-II is supported by DOE, BES-FWP-PS001. We thank Shirish Chodankar, Lin Yang, Derrick Lin, Suzette Pabit and Josue San Emeterio for the assistance at the LiX beamline. This research was also conducted with support from the Cornell University Center for Advanced Computing, which receives funding from Cornell University, the National Science Foundation, and members of its Partner Program. The authors are grateful to Seyed-Fakhreddin Torabi for sharing the MALAT1 samples and David H. Mathews for providing the sequence of the ssRNA. 

\bibliographystyle{unsrt}
\bibliography{references}

\renewcommand\thefigure{S\arabic{figure}}    
\setcounter{figure}{0} 
\newpage

\begin{figure}[h]
\centering
\includegraphics[width=\textwidth]{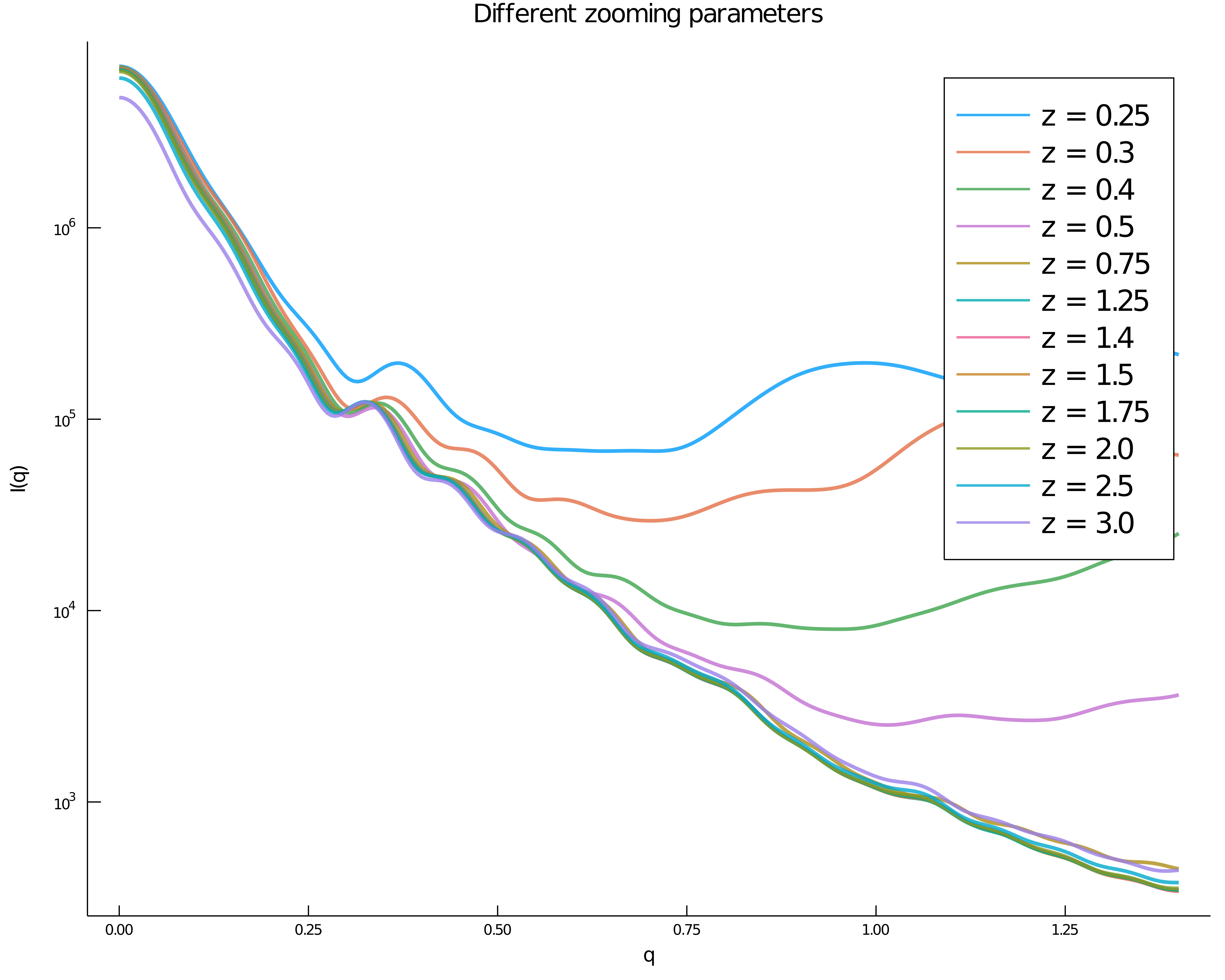}
\caption{The effect of different voxel splitting sizes on the SWAXS profile. The splitting parameter, $z$, has minimal effect on the SWAXS profile when $z > 0.75$. Note that in the \textit{FM} algorithm, it marches to higher $q$ and $z > 1.0$ always.}
\label{fig:S1}
\end{figure}

\newpage
\begin{figure}[h]
\centering
\includegraphics[width=\textwidth]{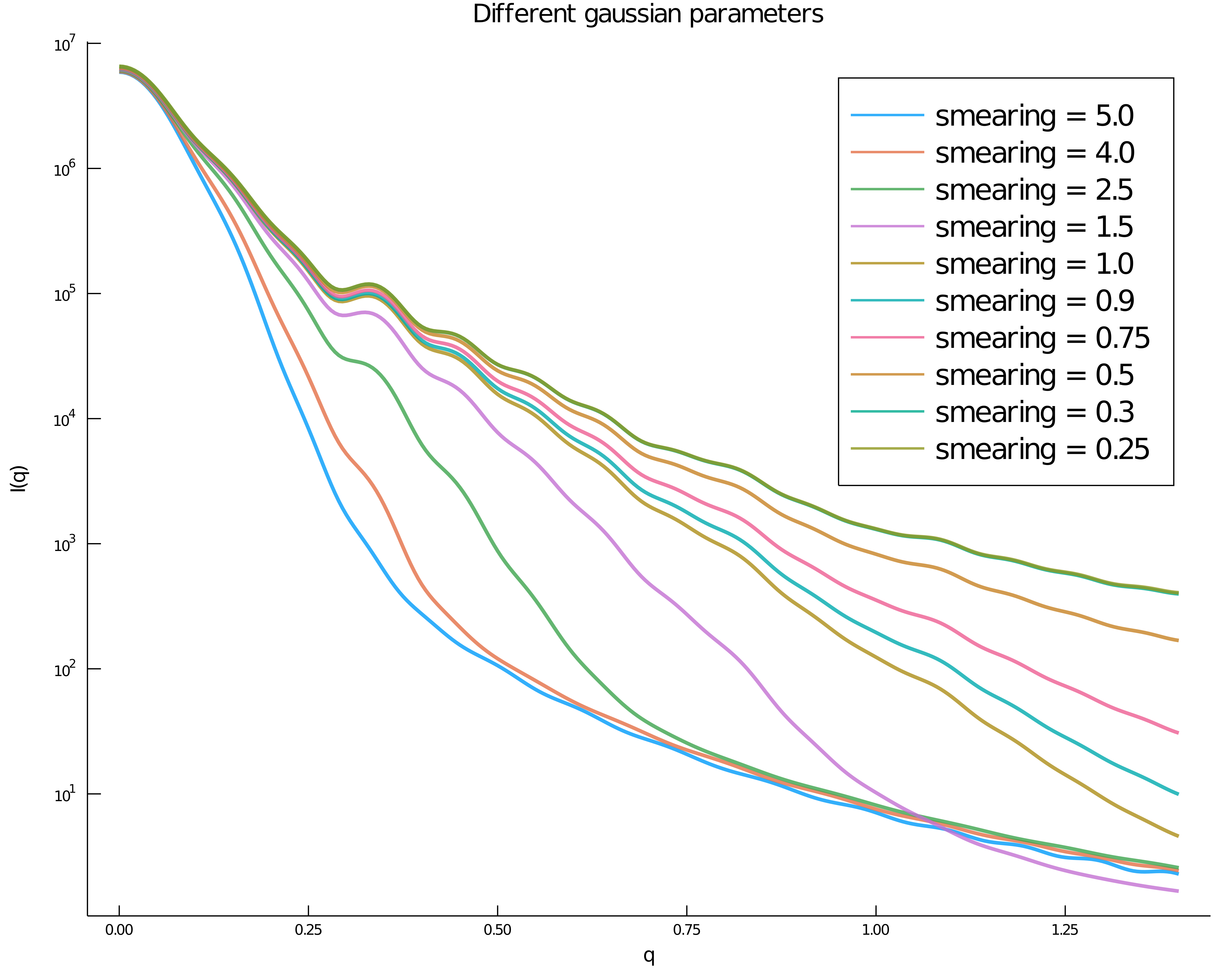}
\caption{The effect of electron density smearing on the SWAXS profile. The smearing parameter, $\sigma$, has minimal effects on the SWAXS profile when $\sigma < 1.0$. Note that in the FM algorithm, $0.3 < \sigma < 1.0$ always. }
\label{fig:S2}
\end{figure}

\newpage
\begin{figure}[h]
\centering
\includegraphics[width=\textwidth]{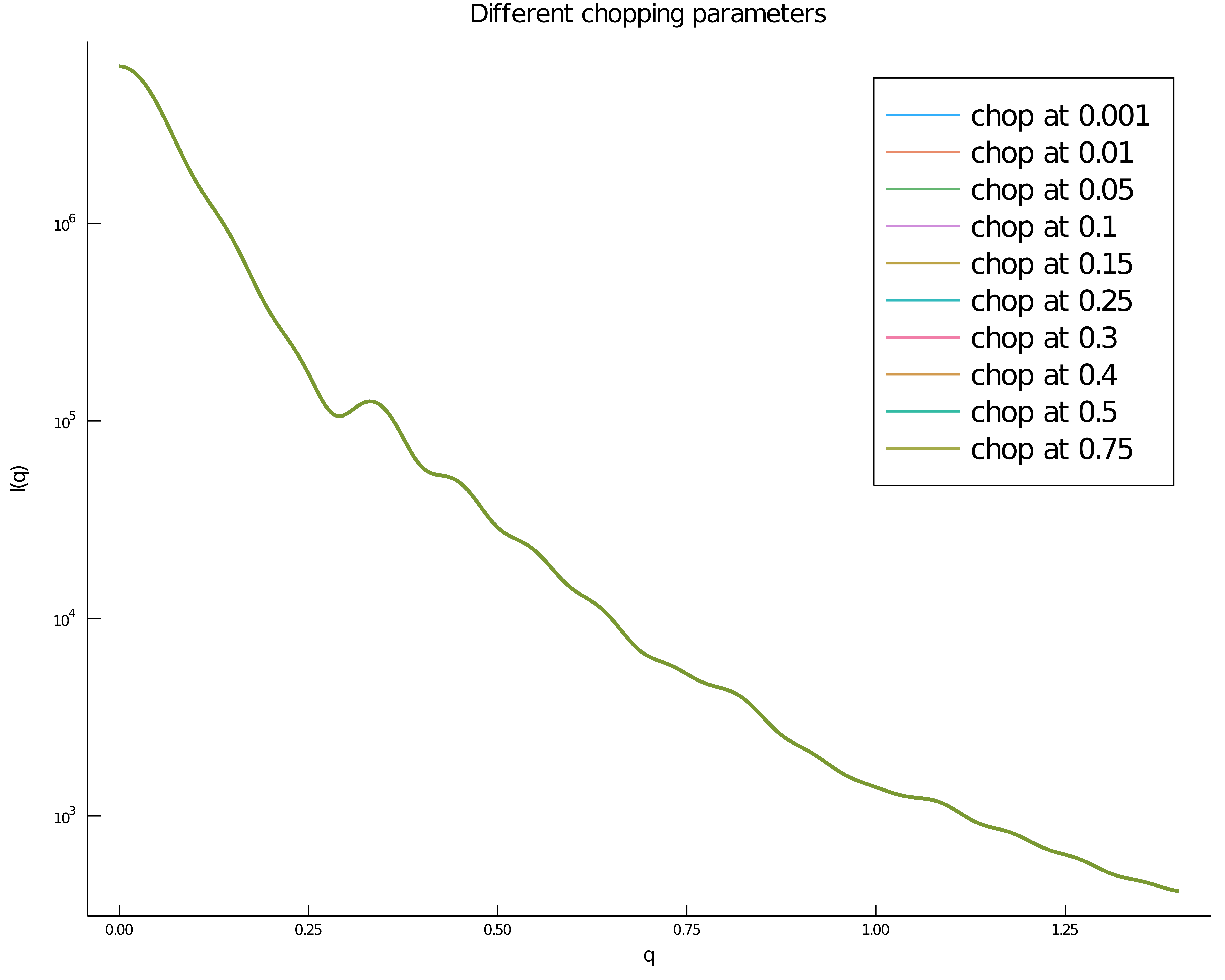}
\caption{The effect of different widths of solvent layers on the SWAXS profile. The chop parameter has minimal effect on the SWAXS profile due to our use of uniform implicit solvent model. Note that in the FM algorithm, we used the largest width of solvent layer corresponding to chop parameter of $0.001$. }
\label{fig:S3}
\end{figure}




\end{document}